\documentclass[]{interact}

\usepackage{epstopdf}
\usepackage[caption=false]{subfig}
\usepackage{color}

\theoremstyle{plain}

\theoremstyle{definition}

\theoremstyle{remark}

\begin{document}

\articletype{}

\title{Expanding the deep-learning model to diagnosis LVNC: Limitations and trade-offs}

\author{
\name{G. Bernabé\textsuperscript{a}\thanks{CONTACT G. Bernabé Email: gbernabe@um.es},  P. González-Férez\textsuperscript{a}, J. M. García\textsuperscript{a}, G. Casas\textsuperscript{b} and J. González-Carrillo\textsuperscript{c}}
\affil{\textsuperscript{a}Computer Engineering Department, University of Murcia, Murcia, Spain; \textsuperscript{b}Hospital Universitari Vall d'Hbron, Barcelona, Spain;
\textsuperscript{c}Hospital Virgen de la Arrixaca, Murcia, Spain}
}

\maketitle

\begin{abstract}
Hyper-trabeculation or non-compaction in the left ventricle of the myocardium (LVNC) is a recently classified form of cardiomyopathy. Several methods have been proposed to quantify the trabeculae accurately in the left ventricle, but there is no general agreement in the medical community to use a particular approach. In previous work, we proposed DL-LVTQ, a deep learning approach for left ventricular trabecular quantification based on a U-Net CNN architecture. DL-LVTQ was an automatic diagnosis tool developed from a dataset of patients with the same cardiomyopathy (hypertrophic cardiomyopathy). 

In this work, we have extended and adapted DL-LVTQ to cope with patients with different cardiomyopathies. The dataset consists of up $379$ patients in three groups with different particularities and cardiomyopathies. Patient images were taken from different scanners and hospitals. We have modified and adapted the U-Net convolutional neural network to account for the different particularities of a heterogeneous group of patients with various unclassifiable or mixed and inherited cardiomyopathies. 

The inclusion of new groups of patients has increased the accuracy, specificity and kappa values while maintaining the sensitivity of the automatic deep learning method proposed. Therefore, a better-prepared diagnosis tool is ready for various cardiomyopathies with different characteristics. 
Cardiologists have considered that $98.9\%$ of the evaluated outputs are verified clinically for diagnosis. Therefore, the high precision to segment the different cardiac structures allows us to make a robust diagnostic system objective and faster, decreasing human error and time spent. 

\end{abstract}


\section{Introduction}

According to the World Health Organization reports~\cite{CVDsWHO,Namara-19}, cardiovascular diseases are one of the leading causes of death globally, causing about 32\% of all deaths worldwide. Among cardiovascular diseases, Left Ventricular Non-Compaction (LVNC) is a recently classified form of cardiomyopathy characterized by abnormal trabeculations or non-compacted tissue in the left ventricle cavity \cite{towbin2015lvnc}, that can be found in association with other cardiomyopathies \cite{doi:10.1177/1753944713504639, ARBUSTINI20141840, associatedLVNC}. 

Several methods based on magnetic resonance imaging (MRI) have been proposed to accurately quantify the trabeculae in the left ventricle (LV) of the myocardium \cite{Jacquier-10,Captur-13, Captur-14,Choi2016,BERNABEGARCIA2017405}. However, there is a disagreement in the cardiology community to determine a universal standard accompanied by excessive time to obtain the surfaces manually and the subjectivity of the cardiologist to carry it out. 
An entirely understandable clarifying measure to diagnose this cardiomyopathy is to calculate the percentage of the trabecular volume to the total volume of the non-compacted wall of the left ventricle ($VT\%$).

%
In the last decade, Deep Neural Networks (DNNs) have become extremely popular and 
one of the main methods of modern Artificial Intelligence (AI) \cite{Survey-AI}.
DNNs are widely used in many domains and the number of applications that use them has significantly increased. DNNs are employed in almost all scientific fields, such as image recognition, speech recognition, or autonomous vehicles, and other medical areas, such as cancer detection or protein folding prediction. In most of these domains, DNNs can outperform human capabilities thanks to their ability to perform high-level abstractions from large datasets \cite{Survey-AI}.

In fact, several works have recently proposed an automatic solution based on deep learning (DL) techniques to determine the left ventricle volume through MRI \cite{litjens2017survey, cardiacReview}, a crucial issue in assessing cardiac diseases. By using high-performance computing,
recent publications exploit various possibilities of deep learning to segment the left and the right ventricle \cite{PEREZPELEGRI2021106275, Penso2021, Li2021}.
Moreover, Bartolini \emph{et al.}~\cite{Bartoli2020} has proposed a deep learning framework to estimate LVNC based on a DenseNet convolutional neural network (CNN) architecture; however, the trabeculae area is not assessed with enough precision.

Our research group has worked on this topic intensively. In previous works, we proposed a semi-automatic software (QLVTHC) \cite{amia2020, jcm21-Bernabe} and an automatic tool (SOST) \cite{gbernabe} that delineate segmentations of the endocardium border and trabecular masses. Both proposals were based on traditional computer vision techniques. Lastly, we have proposed an automatic tool called \textbf{D}eep \textbf{L}earning for \textbf{L}eft \textbf{V}entricular \textbf{T}rabecular \textbf{Q}uantification (DL-LVTQ) based on the usage of Convolutional Neural Networks (CNNs)~\cite{RODRIGUEZDEVERA2022106548}. In particular, this proposal uses as CNN model the well-known U-Net architecture~\cite{ronneberger_fischer_brox_2015} that provides fast and precise segmentation of images. DL-LVTQ has been able to accurately segment the endocardium border and trabeculae and estimate the
level of hyper-trabeculation to determine the diagnosis of LVNC. This work has been carried out on 277 patients with hypertrophic cardiomyopathy (HCM). 

In this paper, we ask whether this model could be extended to determine the quantification of trabeculae and the diagnosis of LVNC in patients with different types of cardiomyopathies (or multiples cardiomyopathies), which images were taken from different scanners and hospitals. 

The main contributions of this paper are the following:

\begin{itemize}
\item Show in the LVNC diagnosis case the limitations of deep learning models and explain its causes.
\item Modify and adapt our previous U-Net CNN architecture to train DL-LVTQ with images collected at different hospitals and several scanners with various cardiomyopathies.
\item Validate the new proposal by statistical methods and a new group of medical doctors that endorse our experimental results.
\end{itemize}

\section{Materials and Methods}\label{sec:methods}

\subsection{Hyper-trabeculation quantification in the left ventricle}
\label{sec:ht_estimation}

Due to the continuing discrepancy in the medical community in determining the level of left ventricular trabeculation and the diagnosis of LVNC, we made a semi-automatic proposal called QLVTHC \cite{BERNABEGARCIA2017405}, and an automatic tool called SOST \cite{gbernabe} to help them. 

\begin{figure}[!htbp]
    \centering
    \includegraphics[width=2.2in]{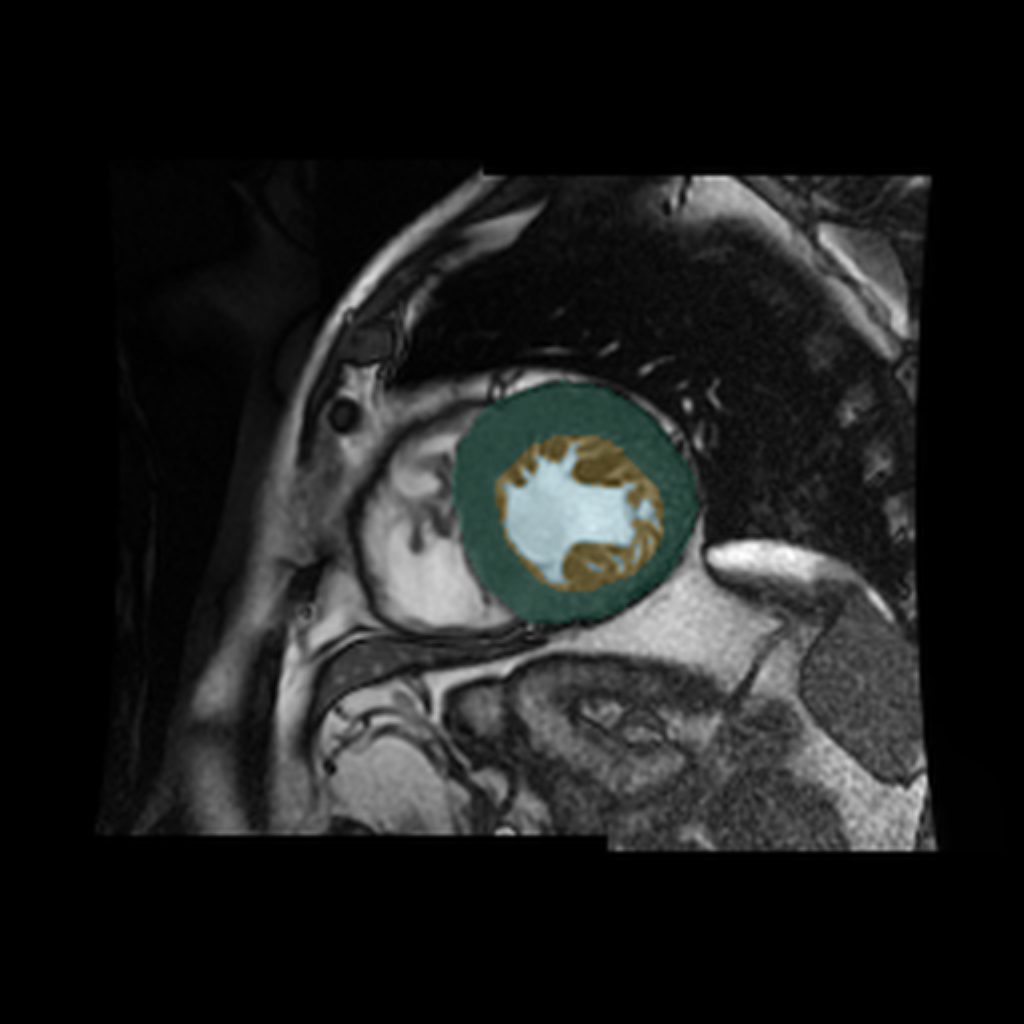}
    \caption{Segmentation of left ventricle, highlighting the external layer (green), the left ventricle cavity (blue) and the trabecular zone or non-compacted area (yellow).}
    \label{fig:sample_segmentation}
\end{figure}

In those proposals, the first stage comprises segmenting the different contours of short-axis cardiac MRI. Consequently, our proposal can detect three different shapes of the left ventricle: the compacted external layer, the trabecular zone and the left ventricle cavity. Figure~\ref{fig:sample_segmentation} shows an example with the three contours detected. As a second step, the area of the trabecular zone and the compacted zone are calculated. Then we obtain the proportion of the trabecular area to the size of the compacted zone for all slices of a patient. Once all areas of a patient are calculated, the trabecular and compacted volumes are obtained by aggregating the information of all slices.

In this way, the percentage of the trabecular volume to the total volume of the external wall of the left ventricle ($VT\%$) is evaluated following Equation~\ref{eq:percent_expression}: 

\begin{equation}\label{eq:percent_expression}
VT\% = 100 \cdot \frac{\textnormal {\textit{Trab. volume}}}{\textnormal {\textit{Trab. volume}}+\textnormal {\textit{Compacted volume}}} \quad [\%]
\end{equation}

\textcolor{black}{The computation of the $VT\%$ has been performed by adding up the areas obtained for each of the individual slices, as was already computed previously in \cite{gbernabe}, taking into account that the slice thickness and interslice spacing are the same for all the slices of a specific patient.}


High values of this metric have been used as a cut-off point or indicator of LVNC \cite{Jacquier-10, BERNABEGARCIA2017405, gbernabe}, enabling a diagnosis. In fact, in the tool QLVTHC, we proposed a validated threshold of $27.4\%$ to differentiate between patients with LVNC and healthy patients~\cite{BERNABEGARCIA2017405}.

\subsection{The DL-LVTQ approach}
\label{sec:nna}

In our previous work \cite{RODRIGUEZDEVERA2022106548}, we have selected a \textcolor{black}{2D} U-Net-like CNN
based on the well-known U-Net \cite{ronneberger_fischer_brox_2015}, from among other networks \cite{lvncDL} because it provides the best performance to the segmentation of the three contours in the left ventricle.

A U-Net architecture contains a symmetric encoder and decoder pathway, with skip connections between the corresponding layers.
The context information is encoded in the down-sampling path and transferred with skip connections to the up-sampling path to decoding feature-map 
and the segmentation maps.

This U-Net architecture was trained on \textcolor{black}{2D} short-axis magnetic resonance imaging to segment the left ventricle's internal cavity, external wall and trabecular tissue, obtaining the following numbers for 25 test images: the average and standard deviation of the Dice coefficient for the internal cavity, external wall and trabeculae were $0.96\pm0.00$, $0.89\pm0.00$ and $0.84\pm0.00$, respectively. Besides, to validate the diagnosis, the outputs generated automatically by DL-LVTQ were visually graded by two cardiologists according to Gibson's scale \cite{gibson_spann_woolley_2004}, \textcolor{black}{\cite{Zaid-08, ICCS15-gbernabe}}, over a set of $25$ patients and $99.5\%$ of the slices were determined without diagnostically relevant issues, \textcolor{black}{that is, the segmentation in internal cavity, external wall and trabecular tissue was visually correct. Note that using Gibson’s scale, the outputs generated 
were visually graded 
from 1 to 5. A value higher than or equal to 4.0 indicates that the segmentation does not represent diagnostically significant differences.}

This paper is the result of the following question: Can the DL-LVTQ approach be used for diagnosing new images not seen before? The problem with deep learning-based techniques is that sometimes they have difficulties extracting the main features of the images. Then, these approaches obtain good accuracy for the learning images, but their scores drop for test images (the over-fitting problem). In the former paper \cite{RODRIGUEZDEVERA2022106548} we did a five-fold cross-validation process to ensure the robustness of the results, splitting into five equal-sized folds the image dataset. One of these folds was used as a test set, and the model was trained using the other four-folds (split into training-validation sets with 80\%–20\%). \textcolor{black}{It is important to note that the training, validation and test sets were defined based on individual patients.} 


However, we were concerned regarding the dataset, as all the patients (277, which has a mean of 7 images per patient) suffered from the same cardiomyopathy. Therefore, we extended our original working dataset into three populations: the original P group (augmented with \textcolor{black}{16} new patients, \textcolor{black}{therefore the P group has 293~patients}), the X group and the H group (see Section 2.4 
for more details on these populations). Table~\ref{tab:inference1} gives our new results on diagnosing (inferencing) on a different set of populations P, X and H, showing the mean and standard deviation of the Dice coefficients for the three contours. The model trained individually on the set P inferences patients of populations X and H with a significant drop in the Dice coefficients reaching $0.70$ in the trabecular zone for patients belonging to H, causing difficulties in reconstructing some slices. The average Dice value for the different contours decreases to $0.84$ and $0.80$ for patients of sets X and H, respectively. Therefore, the neural network model training with images from the P populations do not obtain a good generalization for patients from other populations. 

\begin{table*}[h]
\centering
\begin{tabular}
    {@{}
    l
    c
    c
    c
    c
    @{}}
\toprule
{Population} & \multicolumn{1}{c}{Dice CEL} & \multicolumn{1}{c}{Dice LVC} & \multicolumn{1}{c}{Dice TZ} & Average Dice  \\ \hline \midrule
P on P          & $0.89 \pm 0.10$ &  $0.94 \pm 0.10$ & $0.83 \pm 0.15$ & $0.89 \pm 0.09$  \\
X on P          & $0.82 \pm 0.16$ &  $0.92 \pm 0.15$ & $0.78 \pm 0.19$ & $0.84 \pm 0.15$  \\
H on P          & $0.83 \pm 0.13$ &  $0.88 \pm 0.13$ & $0.70 \pm 0.21$ & $0.80 \pm 0.13$  \\
\hline
\bottomrule
\end{tabular}
\caption{Mean \%($\pm$ standard deviation across the 5 folds) 
of the Dice coefficient for the compacted external layer (CEL), the left ventricle cavity (LVC), the trabecular zone (TZ) and the average of the three contours to inference on different populations}
\label{tab:inference1}
\end{table*}

\subsection{Neural Network architecture}
\label{sec:nna2}

After achieving those results, we looked for reasons that could explain them. 

The first point was the image quality. Input images that conform to the ground-truth delineation were stored at hospitals with low \textcolor{black}{size} ($128\times 128$ \textcolor{black}{pixels}, $92\times 92$ \textcolor{black}{pixels}, and sometimes could decrease to $64\times 64$ \textcolor{black}{pixels}) to save storage space on their hard disks. We realized that \textcolor{black}{image size} is dragging a fundamental problem that limits the perfect reconstruction of the output images. Trying to overcome this issue, we tuned the QLVTHC tool to generate an output image \textcolor{black}{size} of $512\times 512$ \textcolor{black}{pixels} to avoid an excessive loss of ground-truth information.

Afterward, we elaborated on the best neural network architecture to cope with this problem. Based on our previous research on evaluating different networks \cite{lvncDL}, we decided to continue using a U-Net-like network. Then, we have reconfigured the U-Net in the same way as in \cite{RODRIGUEZDEVERA2022106548}, but the \textcolor{black}{image size} of the single-channel inputs is $512\times 512$ \textcolor{black}{pixels} (to match the QLVTHC tool output). Therefore, the number of steps to encode and decode has been increased adequately.

In each step of the down-sampling path (encode phase), which is formed by $7$ levels, a convolutional block composed of two $3\times 3$ convolutions with batch normalization and activations with the rectified linear unit (ReLU) is followed by a max-pooling operation with size $2\times 2$ and stride of 2. The number of feature maps is doubled in each down-sampling step.
A convolution block is used in the bottom layer to link the down-sampling to the up-sampling path, and a bilinear interpolation is implemented in the up-sampling path instead of up-convolution (used in the original U-Net design). Identically to the encoding, in the decode phase (up-sampling path), a series of bilinear interpolation and convolutional blocks are repeated to reach the final convolution layer. 

\begin{figure*}[!htbp]
    \centering
    \includegraphics[width=0.99\textwidth]{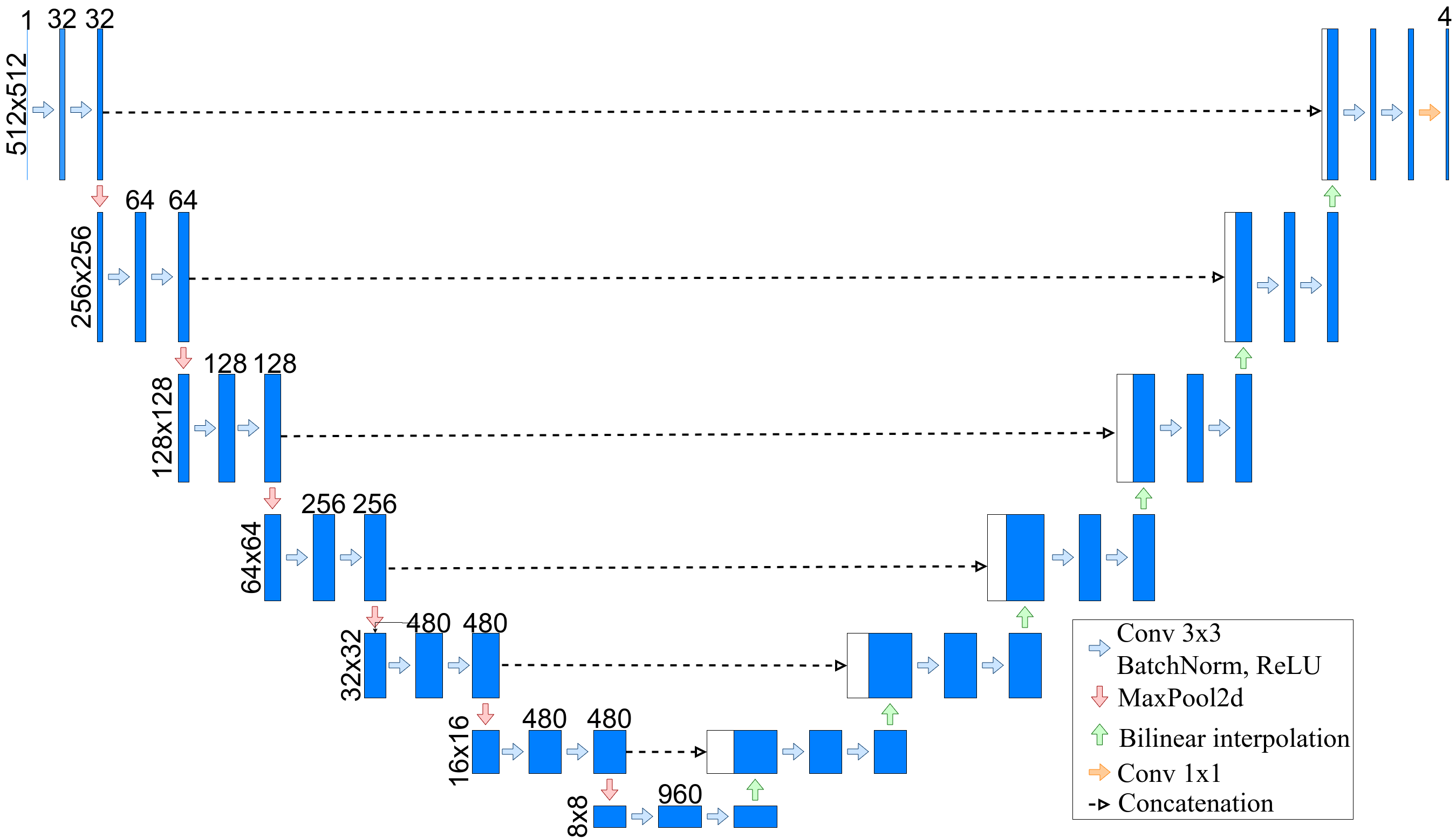}
    \caption{Architecture of the U-Net segmentation network in this research. Each blue box corresponds to a multi-channel feature map.}
    \label{fig:architecture}
\end{figure*}

The final convolution layer includes a softmax activation function to obtain the output segmentation maps in four classes: the left ventricle cavity, the external layer, the trabecular zone and the background. The final segmentation maps also have a \textcolor{black}{size} of $512\times 512$ \textcolor{black}{pixels}. Figure~\ref{fig:architecture} shows the architecture diagram used in our research.

For simplicity, we apply a 2D segmentation to the input slices of the network. We apply Z-score standardization, data augmentation, and a linear combination as a loss function of two components: Lovász-Softmax loss \cite{berman_triki_blaschko_2018}, and weighted binary cross-entropy loss and the \textcolor{black}{Rectified Adam} (RAdam) optimizer \cite{liu2020variance} and a 5-fold cross-validation based on a threshold of 27.4\% previously described in Section 2.1, 
in the same way as in our previous work \cite{RODRIGUEZDEVERA2022106548}.

\subsection{Populations}
\label{sec:populations}

We have compounded a dataset consisting of three groups of patients:

\begin{enumerate}
    
    \item The set labeled P has $293$ patients diagnosed with hypertrophic cardiomyopathy.
    
    
    \item The set labeled X has $58$, unclassifiable patients 
    with different cardiomyopathies diagnosed as non-compacted cardiomyopathy, RV (right ventricular) or LV (left ventricular) arrhythmogenic cardiomyopathy, DCM (dilated cardiomyopathy), HCM (Hypertrophic Cardiomyopathy), unclassifiable or mixed cardiomyopathies, and inherited cardiomyopathies.
    
    \item The set labeled H has $28$ patients 
    with previously diagnosed LVNC cardiomyopathy meeting Petersen’s criteria \cite{Petersen-05}.

\end{enumerate}

The population has been selected from an inherited cardiomyopathy clinic.
The MRI with good quality in the first test has been acquired in the short-axis. 


The MRI images are obtained at three different hospitals: Virgen de la Arrixaca of Murcia (HVAM), Mesa del Castillo of Murcia (HMCM), and Universitari Vall d’Hebron of Barcelona (HUVHB). HVAM manages two 1.5 T scanners, Philips and General Electric, with acquisition matrices of $256\times256$ \textcolor{black}{pixels} and $224\times 224$ \textcolor{black}{pixels} and pixel spacing of
$1.5\times1.5\times0.8mm$ and $1.75\times1.75\times0.8mm$, respectively. HMCM possesses the same General Electric model scanner as HVAM. HUVHB has a 1.5 scanner Avanto of Siemens, where the acquisition matrix is $224 \times 224$ \textcolor{black}{pixels}. The LV function is determined with balanced steady-state free precision (b-SSFP) sequences, where the repetition interval is established to $3.8$ ms for HMCM and HUVHB, whereas HVAM uses $3.3$ ms. Other parameters like echo time, flip angle, echo train length, slice thickness, slice gap and phases are fixed to $1.7$~ms, $60^o$, $23$, $8$~mm, $2$~mm and $20$ phases for all scanners. All patients have been monitored in apnea, in synchronization with the ECG and without a contrast agent.

In short, the final dataset comprises $3044$ slices of $379$ patients. The number of patients with LVNC is $223$. The number of slices per patient ranges from 1 to 14; the median and the mean are seven slices per patient.

\section{Results and discussion} 
\label{sec:results}


\subsection{Training with all the populations} 
\label{subsec:training}

We started by training our modified neural network described in Section 2.3 
with mixed images from different populations.  
Table~\ref{tab:inference2} describe the obtained \textcolor{black}{inference} results with training with images from the P and X populations (P+X in the Table) and when training with images from all populations (P+X+H in the Table). We report the mean and standard deviation of the Dice coefficients for the three contours and the average Dice, split by patients belonging to populations P, X, and H. 

\begin{table*}[h]
\centering
\begin{tabular}
    {@{}
    l
    c
    c
    c
    c
    @{}}
\toprule
{Population} & \multicolumn{1}{c}{Dice CEL} & \multicolumn{1}{c}{Dice LVC} & \multicolumn{1}{c}{Dice TZ} & Average Dice  \\ \hline \midrule
P on P+X        & $0.89 \pm 0.10$ &  $0.94 \pm 0.10$ & $0.84 \pm 0.15$ & $0.89 \pm 0.09$  \\
X on P+X        & $0.83 \pm 0.17$ &  $0.93 \pm 0.16$ & $0.79 \pm 0.21$ & $0.85 \pm 0.16$  \\
H on P+X        & $0.84 \pm 0.12$ &  $0.90 \pm 0.12$ & $0.73 \pm 0.20$ & $0.82 \pm 0.13$  \\
\hline
P on P+X+H      & $0.89 \pm 0.09$ &  $0.94 \pm 0.09$ & $0.84 \pm 0.14$ & $0.89 \pm 0.09$  \\
X on P+X+H      & $0.84 \pm 0.14$ &  $0.93 \pm 0.14$ & $0.80 \pm 0.18$ & $0.86 \pm 0.13$  \\
H on P+X+H      & $0.86 \pm 0.09$ &  $0.92 \pm 0.10$ & $0.79 \pm 0.16$ & $0.86 \pm 0.09$  \\
\hline
\bottomrule
\end{tabular}
\caption{Mean \%($\pm$ standard deviation across the five folds) 
of the Dice coefficient for the compacted external layer (CEL), the left ventricle cavity (LVC), the trabecular zone (TZ), and the average of the three contours to inference on different populations}
\label{tab:inference2}
\end{table*}

As we can see, when the network model is trained with P and X populations (P+X), \textcolor{black}{the Dice coefficients improve concerning those obtained with population P (shown in Table~\ref{tab:inference1})}. However, inferences for patients in the H group remain a bit lower. For example, the Dice coefficient of the trabecular zone and the average Dice value are fixed at $0.73$ and $0.82$, respectively. Finally, when the network model is trained with P, X, and H populations (P+X+H), the results are much better, reaching an average Dice value of $0.86$ to infer both a patient belongs to X or H. Therefore, this whole model, in which several patients with different characteristics of X and H are added to the initial P population, helps to solve and reinforce the model,  making it more robust and prepared to infer patients with distinct heart diseases from different hospitals.



Table~\ref{tab:segmentation_cv5} complete the previous results from our modified neural network. This Table reports \textcolor{black}{the training results} for the mean and standard deviation of the Dice coefficients for the compacted external layer, the trabecular zone, and the left ventricle cavity when using only images from the P population (P), an image mixed from the P and X population (P+X) and, finally, an image combined from the P, X and H population (P+X+H). As we can see, there is high accuracy in detecting the compacted external layer and the left ventricle cavity, keeping very close values even though new sets of patients, such as X and H, are included. This result shows that it is feasible to increase the dataset so that the network is trained for a broader spectrum of patients with distinct features. It is acquired with various machines and in separate hospitals. 
In addition, it is important to note that the standard deviation is very low or close to zero, varying very slightly as new sets of patients are added, showing that the results are pretty robust. 
\begin{table*}[h]
\centering
\begin{tabular}
    {@{}
    l
    c
    c
    c
    c
    @{}}
\toprule
{Population} & \multicolumn{1}{c}{Dice CEL} & \multicolumn{1}{c}{Dice LVC} & \multicolumn{1}{c}{Dice TZ} & Average Dice  \\ \hline \midrule
P               & $0.88 \pm 0.00$ &  $0.95 \pm 0.00$ & $0.83 \pm 0.00$ & $0.89 \pm 0.01$  \\
P $+$ X         & $0.87 \pm 0.02$ &  $0.95 \pm 0.01$ & $0.83 \pm 0.02$ & $0.88 \pm 0.01$  \\
P $+$ X $+$ H   & $0.86 \pm 0.02$ &  $0.94 \pm 0.02$ & $0.82 \pm 0.03$ & $0.88 \pm 0.02$  \\ \hline
\bottomrule
\end{tabular}
\caption{Mean \%($\pm$ standard deviation across the 5 folds) 
of the Dice coefficient for the compacted external layer (CEL), the left ventricle cavity (LVC), the trabecular zone (TZ) and the average of the three contours for different populations.}
\label{tab:segmentation_cv5}
\end{table*}

Regarding the Dice coefficient for the trabecular zone, the obtained values are a bit lesser due to the intrinsic difficulty associated with this zone, formed by several separate parts, an aspect also experienced by cardiologists in determining this controversial area. Therefore, we can claim that the network accurately distinguishes the trabecular zone. Moreover, it is essential to remember that there is a limitation in achieving higher values for the Dice coefficients due to the low \textcolor{black}{image size} at which MRIs are stored in medical centers.








By using the neural network presented in  Section 2.3 
and taking as a reference the values obtained by the semi-automatic QLVTHC proposal~\cite{gbernabe}, we compute the mean error and standard deviation between the $VT\%$, and the values obtained in our modified network model. We report  $5.01\pm  0.19$ \textcolor{black}{$mm^{3}$} for the population P, $5.39\pm 0.90$ \textcolor{black}{$mm^{3}$} for the population P+X and $5.63\pm 1.26$ \textcolor{black}{$mm^{3}$} for the population P+X+H. Therefore, there is no significant increase in committed errors despite expanding the dataset with patients with different cardiomyopathies.

\subsection{Statistical evaluation} 
\label{subsec:training}

From a medical point of view, it is important to evaluate our modified network model statistically. Consequently, the confusion matrices (based on $27.4\%$ threshold validated by QLVTHC) are presented in Table~\ref{fig:conf_matrix_P}, Table~\ref{fig:conf_matrix_P_X} and Table~\ref{fig:conf_matrix_P_X_H} for populations P, P+X and P+X+H, respectively. In addition, Table~\ref{tab:precision} shows the accuracy, sensitivity, specificity and Kappa values. 

\begin{table}[h]
    \centering
    \begin{tabular}{l|c|c|c}
        \multicolumn{1}{c}{}&\multicolumn{2}{c}{Reference diagnosis}&\\
        \cline{2-4}
        \multicolumn{1}{c|}{}&LVNC&No LVNC&\multicolumn{1}{c|}{Total}\\
        \cline{1-4}
        \multicolumn{1}{|c|}{LVNC}&145&22&\multicolumn{1}{c|}{167}\\
        \cline{1-4} 
        \multicolumn{1}{|c|}{No LVNC}&24&123&\multicolumn{1}{c|}{147}\\
        \cline{1-4} 
        \multicolumn{1}{|c}{Total} & \multicolumn{1}{|c|}{$169$} & \multicolumn{1}{c|}{$145$} & \multicolumn{1}{c|}{$314$} \\
        \cline{1-4} 
    
    \end{tabular}
    \caption{Confusion matrix for the population P by a threshold of $27.4\%$.}
    \label{fig:conf_matrix_P}
\end{table}

\begin{table}[h]
    \centering
    \begin{tabular}{l|c|c|c}
        \multicolumn{1}{c}{}&\multicolumn{2}{c}{Reference diagnosis}&\\
        \cline{2-4}
        \multicolumn{1}{c|}{}&LVNC&No LVNC&\multicolumn{1}{c|}{Total}\\
        \cline{1-4}
        \multicolumn{1}{|c|}{LVNC}&193&10&\multicolumn{1}{c|}{203}\\
        \cline{1-4} 
        \multicolumn{1}{|c|}{No LVNC}&39&117&\multicolumn{1}{c|}{156}\\
        \cline{1-4} 
        \multicolumn{1}{|c}{Total} & \multicolumn{1}{|c|}{$232$} & \multicolumn{1}{c|}{$127$} & \multicolumn{1}{c|}{$359$} \\
        \cline{1-4}
    
    \end{tabular}
    \caption{Confusion matrix for the population P+X by a threshold of $27.4\%$}
    \label{fig:conf_matrix_P_X}
\end{table}

\begin{table}[h]
    \centering
    \begin{tabular}{l|c|c|c}
        \multicolumn{1}{c}{}&\multicolumn{2}{c}{Reference diagnosis}&\\
        \cline{2-4}
        \multicolumn{1}{c|}{}&LVNC&No LVNC&\multicolumn{1}{c|}{Total}\\
        \cline{1-4} 
        \multicolumn{1}{|c|}{LVNC}&210&13&\multicolumn{1}{c|}{223}\\
        \cline{1-4} 
        \multicolumn{1}{|c|}{No LVNC}&34&122&\multicolumn{1}{c|}{156}\\
        \cline{1-4} 
        \multicolumn{1}{|c}{Total} & \multicolumn{1}{|c|}{$244$} & \multicolumn{1}{c|}{$135$} & \multicolumn{1}{c|}{$379$} \\
        \cline{1-4}
    \end{tabular}
    \caption{Confusion matrix for the population P+X+H by a threshold of $27.4\%$}
    \label{fig:conf_matrix_P_X_H}
\end{table}

\begin{table*}[h]
\centering
\begin{tabular}
    {@{}
    l
    c
    c
    c
    c
    @{}}
\toprule
{Population} & \multicolumn{1}{c}{Accuracy} & \multicolumn{1}{c}{Sensitivity} & \multicolumn{1}{c}{Specificity} & \multicolumn{1}{c}{Kappa}  \\ \hline \midrule
P               & $0.85$ &  $0.87$   & $0.85$ & $0.71$ \\
P $+$ X         & $0.86$ &  $0.95$   & $0.92$ & $0.72$ \\
P $+$ X $+$ H   & $0.88$ &  $0.94$   & $0.90$ & $0.74$ \\ \hline
\bottomrule
\end{tabular}
\caption{Accuracy, sensitivity, specificity and kappa values for different populations.}
\label{tab:precision}
\end{table*}

As we can see, the accuracy and the kappa of the network are slightly increased, whereas the sensitivity and the specificity are also increased in a more significant way by including the sets X and H. This means we have a better-prepared network for a wide range of cardiomyopathies. This network allows any cardiologist to automatically obtain the inference of any patient without investing a considerable and disproportionate time. Moreover, for the complete dataset, the positive and negative predictive values achieve $0.94$ ($210/223$) and $0.78$ ($122/156$), respectively, improving the detection of LVNC patients and maintaining healthy patients against our previous proposal~\cite{RODRIGUEZDEVERA2022106548}.

\textcolor{black}{Since the cut-off point of $27.4\%$ was validated using the QLVTHC tool, it is possible that a better classification can be achieved by varying the threshold point using DL-QLVT. Hence,it is necessary to calculate receiver operating characteristics (ROC) curve analysis of the $VT\%$ obtained from the modified network to determine that the optimal cut-off point is $27.1\%$. The corresponding confusion matrix can be found in Table~\ref{fig:conf_matrix_P_X_H2}. The area under the ROC curve is $0.94$ ($95\%$ confidence interval, $0.91$--$0.96$), and with a threshold of $27.1\%$, the accuracy, sensitivity, specificity and Kappa values are $0.88$, $0.96$, $0.94$ and $0.75$, respectively. Therefore, the positive and negative predictive values achieve 0.96 ($215/223$) and $0.77$ ($120/156$), respectively, further improving the detection of LVNC patients and very slightly worsening the detection of healthy patients.}

\begin{table}[h]
    \centering
    \begin{tabular}{l|c|c|c}
        \multicolumn{1}{c}{}&\multicolumn{2}{c}{\textcolor{black}{Reference diagnosis}}&\\
        \cline{2-4}
        \multicolumn{1}{c|}{}&\textcolor{black}{LVNC}&\textcolor{black}{No LVNC}&\multicolumn{1}{c|}{\textcolor{black}{Total}}\\
        \cline{1-4}
        \multicolumn{1}{|c|}{\textcolor{black}{LVNC}}&\textcolor{black}{215}&\textcolor{black}{8}&\multicolumn{1}{c|}{\textcolor{black}{223}}\\
        \cline{1-4} 
        \multicolumn{1}{|c|}{\textcolor{black}{No LVNC}}&\textcolor{black}{36}&\textcolor{black}{120}&\multicolumn{1}{c|}{\textcolor{black}{156}}\\
        \cline{1-4} 
        \multicolumn{1}{|c}{\textcolor{black}{Total}} & \multicolumn{1}{|c|}{\textcolor{black}{$215$}} & \multicolumn{1}{c|}{\textcolor{black}{$128$}} & \multicolumn{1}{c|}{\textcolor{black}{$379$}} \\
        \cline{1-4}   
    \end{tabular}
    \caption{\textcolor{black}{Confusion matrix for the population P+X+H by a threshold of $27.1\%$.}}
    \label{fig:conf_matrix_P_X_H2}
\end{table}


\subsection{Medical validation}

Finally, we present an evaluation of the outputs generated automatically by the updated DL-LVTQ. Two cardiologists from two different hospitals carried out this evaluation. We do not intend to introduce a subjective assessment of the images, which contradicts the aim of the proposed automatic tool. Still, it is crucial that cardiologists, different from those who collected the images, can detect any errors.

More specifically, these cardiologists have scored for all slices of patients belonging to population H.
Both cardiologists have determined that a $98.9\%$ of the slices of H-patients are fully valid from a medical point of view (a score higher or equal to 3.5) or without significant differences to make a diagnostic. 

\begin{figure}
\centering
\subfloat[]{%
\resizebox*{4cm}{!}{\includegraphics{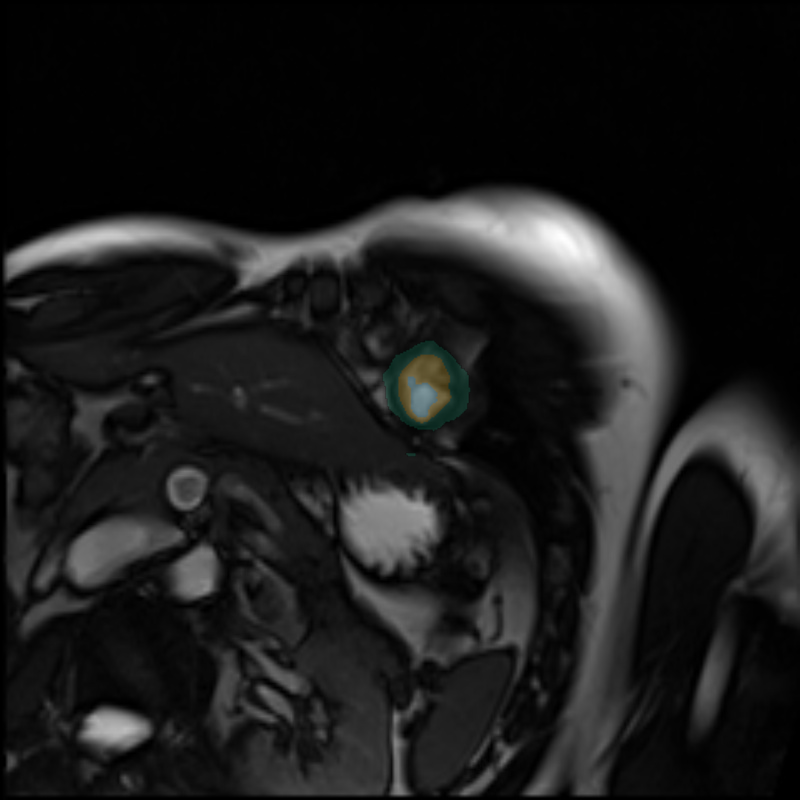}}}\hspace{5pt}
\subfloat[]{%
\resizebox*{4cm}{!}{\includegraphics{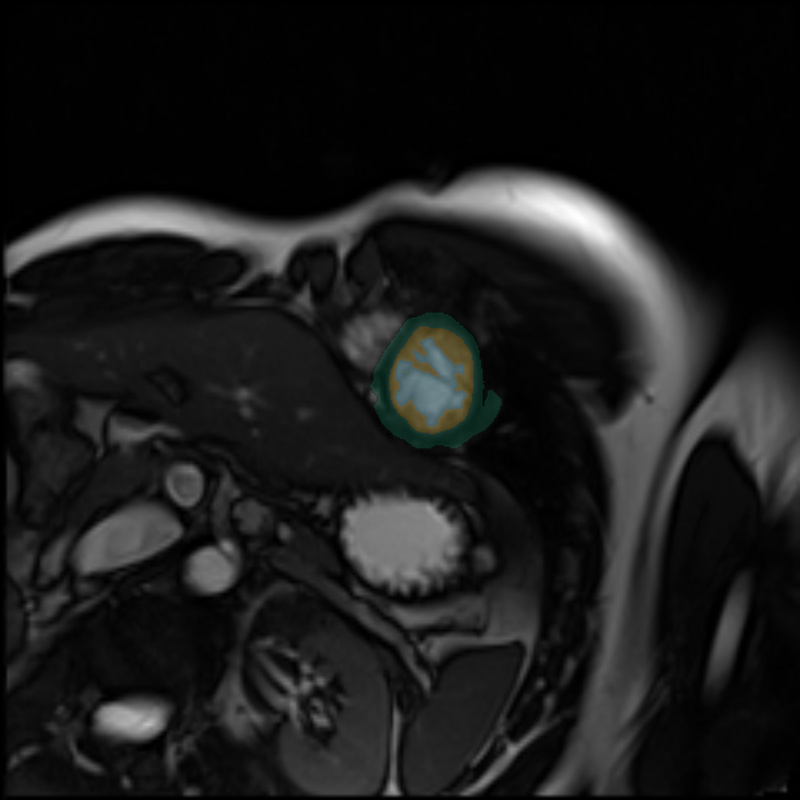}}}\hspace{5pt}
\subfloat[]{%
\resizebox*{4cm}{!}{\includegraphics{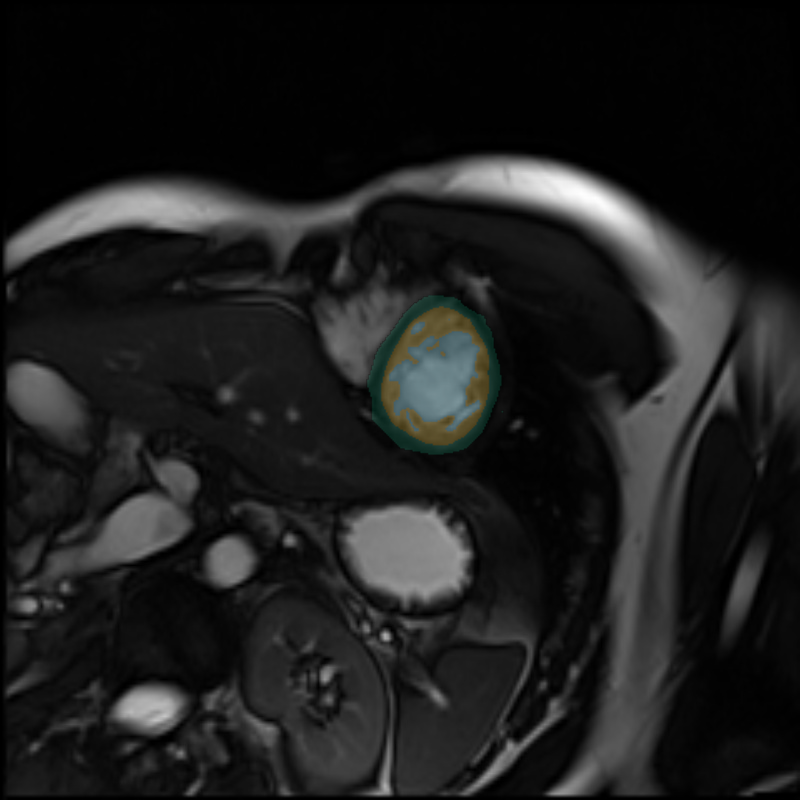}}}\hspace{5pt}
\subfloat[]{%
\resizebox*{4cm}{!}{\includegraphics{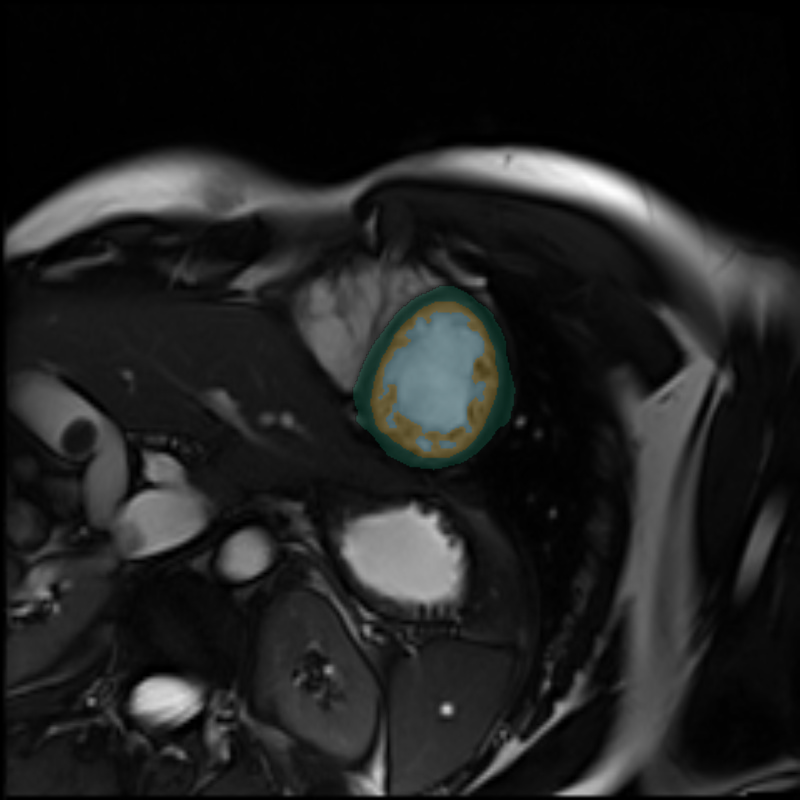}}}\hspace{5pt}
\subfloat[]{%
\resizebox*{4cm}{!}{\includegraphics{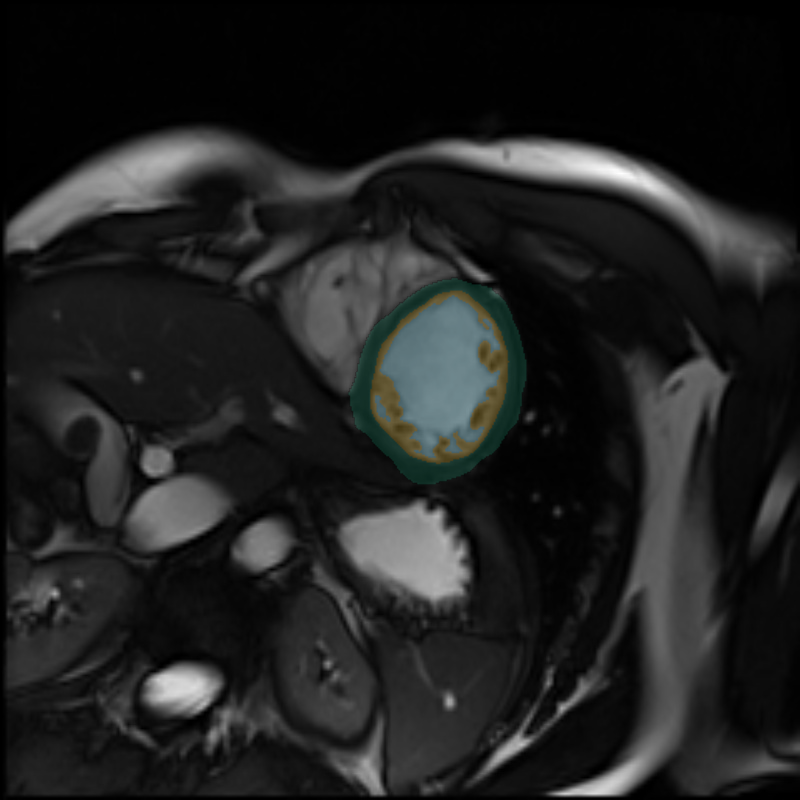}}}\hspace{5pt}
\subfloat[]{%
\resizebox*{4cm}{!}{\includegraphics{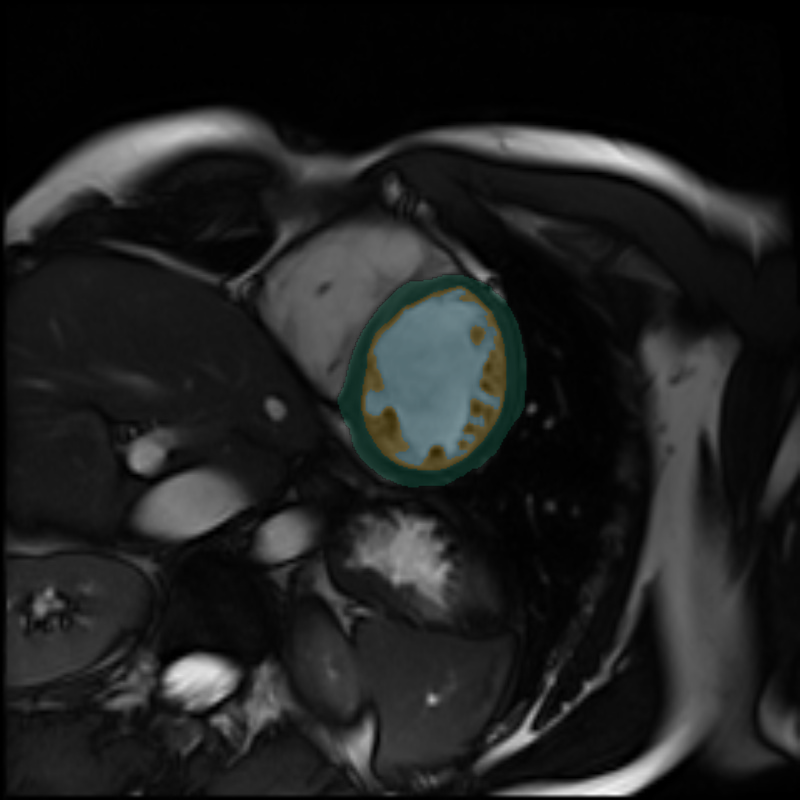}}}\hspace{5pt}
\subfloat[]{%
\resizebox*{4cm}{!}{\includegraphics{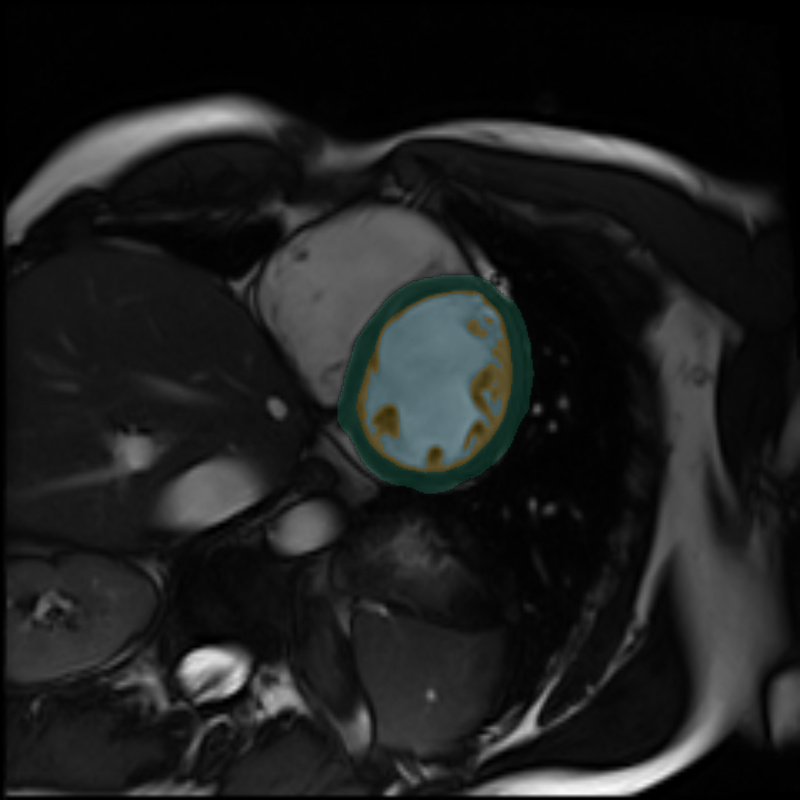}}}\hspace{5pt}
\subfloat[]{%
\resizebox*{4cm}{!}{\includegraphics{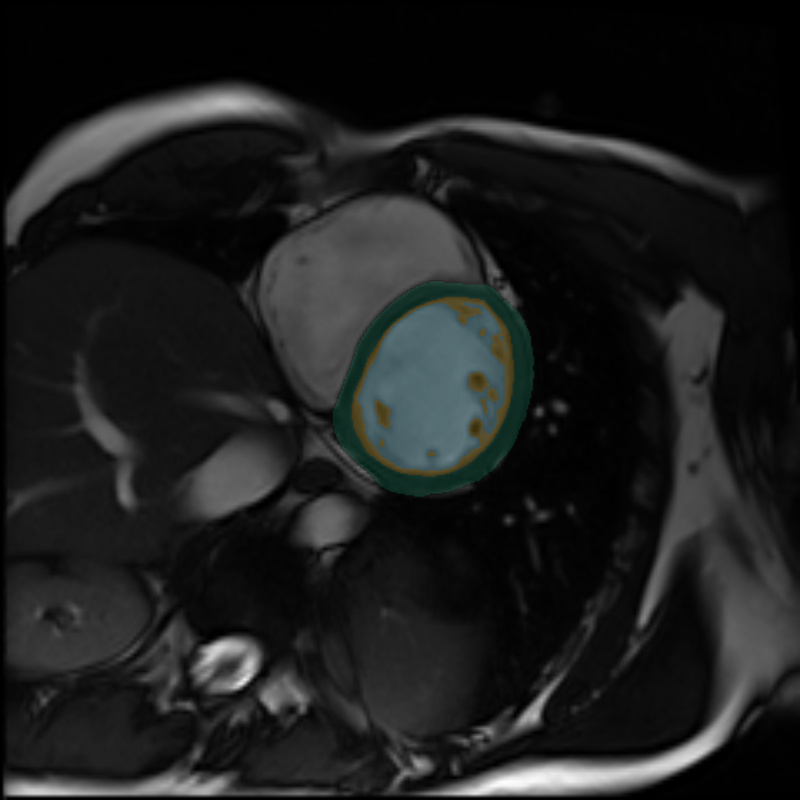}}}\hspace{5pt}
\subfloat[]{%
\resizebox*{4cm}{!}{\includegraphics{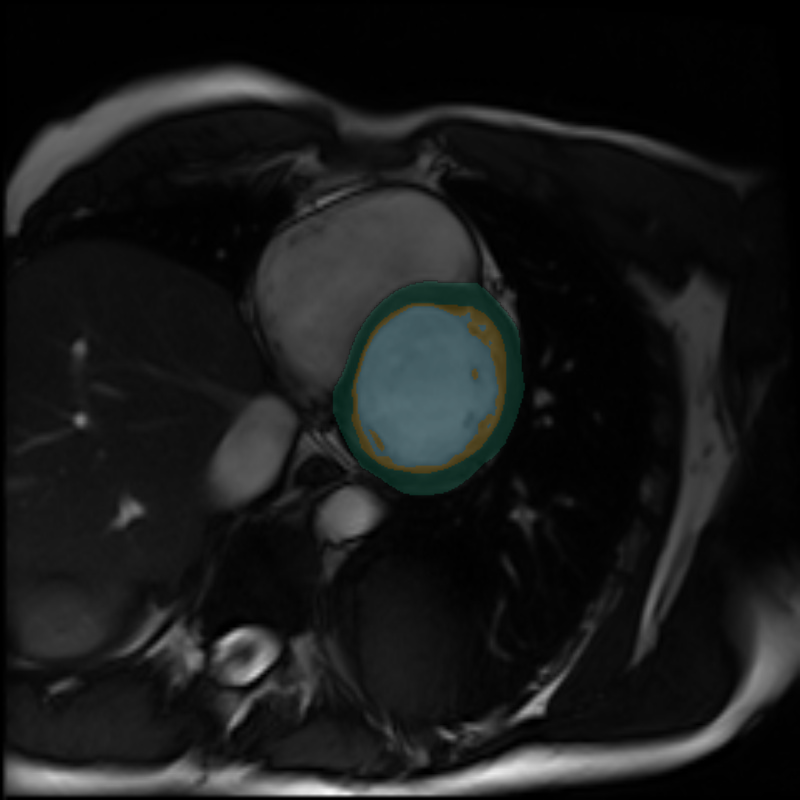}}}
\caption{Output slices for the patient H15. Green indicates the compacted external layer of the left ventricle cavity and yellow the trabecular zone.} \label{fig:outputs_FAAJ}
\end{figure}

For example, Figure~\ref{fig:outputs_FAAJ} show the segmentations (green indicates the compacted external layer of the left ventricle cavity and yellow the trabecular zone) for one patient, where two cardiologists scored a 5 (exact match) or 4.5 on the different slices. 
However, fewer slices of H patients have not been rebuilt correctly, so it could be advisable to increase the number of such patients in the dataset, causing a double benefit that will imply an improvement of the Dice coefficient for the different contours and better image reconstruction.

Regarding inter-observer variability, the mean difference in the scores given by both cardiologists was $0.58\pm 0.51$ ($13.1\%\pm11.8\%$). This value shows a discrepancy due to the subjectivity mentioned above of the different cardiologists from other geographically separated hospitals.
\textcolor{black}{In particular, the discrepancy lies mainly in the slices close to
the apical or basal ends, where the accuracy of the network is lesser due to their tiny sizes in apical-ends slices or the delimitation of borders in the adjacent cavities for the basal-ends slices as it has been reported previously~\cite{RODRIGUEZDEVERA2022106548, 8360453}.}


\section{Conclusions and future work}
\label{sec:conclusions}

In a previous paper, we presented DL-LVTQ \cite{RODRIGUEZDEVERA2022106548}. DL-LVTQ was a deep learning approach for left ventricular trabecular quantification based on a U-Net CNN architecture. DL-LVTQ was ready to process and determine the presence of LVNC in patients with hypertrophic cardiomyopathy. 

 From a medical point of view, the results generated by DL-LVTQ do not present notable differences for making a diagnosis in most cases, which favors inter-hospital collaboration and research in difficult-to-treat patients with numerous or unclassifiable cardiomyopathies.
 
 In this paper, we modify the U-Net and adapt it to process the different particularities of three groups of patients with various unclassifiable or mixed and inherited cardiomyopathies. The images of these new groups of patients were obtained at different hospitals, acquired by several scanners, and conformed to the new dataset. 
 
Our modified network model achieves excellent accuracy in segmenting the endocardium border and trabeculae and outperforms the previous manual, semi-automatic and automatic proposals robustly and quickly. Therefore, cardiologists automatically obtain the inference of any patient without significant effort and personal subjectivity. Moreover, we can claim that the system is prepared and generalized for a heterogeneous group of patients.


\textcolor{black}{To sum up, the automatic diagnosis of LNVC cardiomyopathy by deep learning approaches from MRI images is a complex a complicated process. However, we consider our result worthy and promising. Although there is still a way to cover, the automatic diagnosis will provide cardiologists with an automated, fast and robust system for determining LVNC without spending considerable time, eliminating human error and subjectivity and favoring the evolution of LVNC and other heart diseases. This collaboration among hospitals and research groups will allow the diagnosis of unclassified patients with multiple cardiomyopathies. However, among all parties, this approach will most benefit the patient.}

\section*{Data Availability} 

The dataset used to support the findings of this study was approved by the local ethics committee, and so cannot be made freely available. Requests for access to these data should be made to the corresponding author, Gregorio Bernabé, gbernabe@um.es.

\section*{Conflict of interest}

The authors declare that there is no conflict of interest.

\section*{Acknowledgements}

Grant TED2021-129221B-I00 funded by MCIN/AEI/10.13039/501100011033 and by the “European Union NextGenerationEU/PRTR”.

\bibliographystyle{tfcse}
\bibliography{mybibfile}

\begin{thebibliography}{31}
\providecommand{\natexlab}[1]{#1}
\providecommand{\url}[1]{\normalfont{#1}}
\providecommand{\urlprefix}{Available from: }

\bibitem[Arbustini et~al.(2014)]{ARBUSTINI20141840}
Arbustini~E, Weidemann~F, Hall~JL. 2014. Left ventricular noncompaction: a distinct cardiomyopathy or a trait shared by different cardiac diseases? Journal of the American College of Cardiology. 64(17):1840 -- 1850.

\bibitem[Bartoli et~al.(2020)]{Bartoli2020}
Bartoli~A, Fournel~J, Bentatou~Z, G~GH, Lalande~A, Bernard~M, Boussel~L, Pontana~F, Dacher~J, Ghattas~B, et~al. 2020. Deep learning-based automated segmentation of left ventricular trabeculations and myocardium on cardiac {MR} images: A feasibility study. Radiology Artificial intelligence. 25(3):--.

\bibitem[Berman et~al.(2018)]{berman_triki_blaschko_2018}
Berman~M, Triki~AR, Blaschko~MB. 2018. The {L}ovasz-softmax loss: a tractable surrogate for the optimization of the intersection-over-union measure in neural networks. 2018 IEEE/CVF Conference on Computer Vision and Pattern Recognition{ }:--.

\bibitem[Bernab\'{e} et~al.(2018)]{gbernabe}
Bernab\'{e}~G, Casanova~JD, Cuenca~J, Gonz\'{a}lez-Carrillo~J. 2018. A self-optimized software tool for quantifying the degree of left ventricle hyper-trabeculation. The Journal of Supercomputing. 75(3):1625–1640.

\bibitem[Bernab\'{e} et~al.(2015)]{ICCS15-gbernabe}
Bernab\'{e}~G, Cuenca~J, Gim\'{e}nez~D, L\'{o}pez~de Teruel~PE, Gonz\'{a}lez-Carrillo~J. 2015. A software tool for the automatic quantification of the left ventricle myocardium hyper-trabeculation degree. Procedia Computer Science. 51:610--619.

\bibitem[{Bernabé} et~al.(2020)]{amia2020}
{Bernabé}~G, {Casanova}~JD, {Casas}~G, {González-Carrillo}~J. 2020. A highly accurate method for quantifying lvnc cardiomyophaty. AMIA 2020 Annual Symposium:223 -- 232.

\bibitem[Bernabé et~al.(2021)]{jcm21-Bernabe}
Bernabé~G, Casanova~JD, González-Carrillo~J, Gimeno-Blanes~JR. 2021. Towards an enhanced tool for quantifying the degree of lv hyper-trabeculation. Journal of Clinical Medicine. 10(3):--.

\bibitem[{Bernabé} et~al.(2017)]{BERNABEGARCIA2017405}
{Bernabé}~G, González-Carrillo~J, {Cuenca}~J, {Rodríguez}~D, {Saura}~D, {Gimeno}~JR. 2017. Performance of a new software tool for automatic quantification of left ventricular trabeculations. Revista Española de Cardiología (English Edition). 70(5):405 -- 407.

\bibitem[{Bernard} et~al.(2018)]{8360453}
{Bernard}~O, {Lalande}~A, {Zotti}~C, {Cervenansky}~F, {Yang}~X, {Heng}~PA, {Cetin}~I, {Lekadir}~K, {Camara}~O, {Gonzalez Ballester}~MA, et~al. 2018. Deep learning techniques for automatic mri cardiac multi-structures segmentation and diagnosis: Is the problem solved? IEEE Transactions on Medical Imaging. 37(11):2514--2525.

\bibitem[Biagini et~al.(2006)]{associatedLVNC}
Biagini~E, Ragni~L, Ferlito~M, et~al. 2006. Different types of cardiomyopathy associated with isolated ventricular noncompaction. The American journal of cardiology. 98:821--4.

\bibitem[Captur et~al.(2014)]{Captur-14}
Captur~G, Lopes~LR, Patel~V, Li~C, Bassett~P, Syrris~P, Sado~DM, Maestrini~V, Mohun~TJ, McKenna~WJ, et~al. 2014. Abnormal cardiac formation in hypertrophic cardiomyopathy. Circulation: Cardiovascular Genetics. 7(3):241--248.

\bibitem[Captur et~al.(2013)]{Captur-13}
Captur~G, Muthurangu~V, Cook~C, Flett~AS, Wilson~R, Barison~A, Sado~DM, Anderson~S, McKenna~WJ, Mohun~TJ, et~al. 2013. Quantification of left ventricular trabeculae using fractal analysis. Journal of Cardiovascular Magnetic Resonance. 15(1):36.

\bibitem[Chen et~al.(2020)]{cardiacReview}
Chen~C, Qin~C, Qiu~H, Tarroni~G, Duan~J, Bai~W, Rueckert~D. 2020. Deep learning for cardiac image segmentation: a review. Frontiers in Cardiovascular Medicine. 7:--.

\bibitem[Choi et~al.(2016)]{Choi2016}
Choi~Y, Kim~SM, Lee~SC, Chang~SA, Jang~SY, Choe~YH. 2016. Quantification of left ventricular trabeculae using cardiovascular magnetic resonance for the diagnosis of left ventricular non-compaction: evaluation of trabecular volume and refined semi-quantitative criteria. Journal of cardiovascular magnetic resonance : official journal of the Society for Cardiovascular Magnetic Resonance. 18(1):24--24.

\bibitem[Gibson et~al.(2004)]{gibson_spann_woolley_2004}
Gibson~D, Spann~M, Woolley~S. 2004. A wavelet-based region of interest encoder for the compression of angiogram video sequences. IEEE Transactions on Information Technology in Biomedicine. 8(2):103–113.

\bibitem[Jacquier et~al.(2010)]{Jacquier-10}
Jacquier~A, Thuny~F, Jop~B, Giorgi~R, Cohen~F, Gaubert~JY, Vidal~V, Bartoli~JM, Habib~G, Moulin~G. 2010. Measurement of trabeculated left ventricular mass using cardiac magnetic resonance imaging in the diagnosis of left ventricular non-compaction. European Heart Journal. 31(9):1098--1104.

\bibitem[Li et~al.(2021)]{Li2021}
Li~C, Song~X, Zhao~H, Feng~L, Hu~T, Zhang~Y, Jiang~J, Wang~J, Xiang~J, Sun~Y. 2021. An 8-layer residual {U-Net} with deep supervision for segmentation of the left ventricle in cardiac ct angiography. Computer Methods and Programs in Biomedicine. 200:105876.

\bibitem[Litjens et~al.(2017)]{litjens2017survey}
Litjens~G, Kooi~T, Bejnordi~BE, Setio~AAA, Ciompi~F, Ghafoorian~M, {van der Laak}~JA, {van Ginneken}~B, Sánchez~CI. 2017. A survey on deep learning in medical image analysis. Medical Image Analysis. 42:60--88.

\bibitem[Liu et~al.(2020)]{liu2020variance}
Liu~L, Jiang~H, He~P, Chen~W, Liu~X, Gao~J, Han~J. 2020. On the variance of the adaptive learning rate and beyond. In: 8th International Conference on Learning Representations, {ICLR} 2020, Addis Ababa, Ethiopia, April 26-30, 2020. p.~--.  \urlprefix\url{https://openreview.net/forum?id=rkgz2aEKDr}.

\bibitem[Namara et~al.(2019)]{Namara-19}
Namara~KM, Alzubaidi~H, Jackson~JK. 2019. Cardiovascular disease as a leading cause of death: how are pharmacists getting involved?. Integrated pharmacy research \& practice. 8:1--11.

\bibitem[Penso et~al.(2021)]{Penso2021}
Penso~M, Moccia~S, Scafuri~S, Muscogiuri~G, Pontone~G, Pepi~M, Caiani~E. 2021. Automated left and right ventricular chamber segmentation in cardiac magnetic resonance images using dense fully convolutional neural network. Computer Methods and Programs in Biomedicine. 204:106059.

\bibitem[Petersen et~al.(2005)]{Petersen-05}
Petersen~SE, Selvanayagam~JB, Wiesmann~F, Robson~MD, Francis~JM, Anderson~RH, Watkins~H, Neubauer~S. 2005. {Left Ventricular Non-Compaction: Insights From Cardiovascular Magnetic Resonance Imaging}. {Journal of the American College of Cardiology}. 46(1):101 -- 105.

\bibitem[Pérez-Pelegrí et~al.(2021)]{PEREZPELEGRI2021106275}
Pérez-Pelegrí~M, Monmeneu~JV, López-Lereu~MP, Pérez-Pelegrí~L, Maceira~AM, Bodí~V, Moratal~D. 2021. Automatic left ventricle volume calculation with explainability through a deep learning weak-supervision methodology. Computer Methods and Programs in Biomedicine. 208:106275.  \urlprefix\url{https://www.sciencedirect.com/science/article/pii/S0169260721003497}.

\bibitem[Rodr{\'i}guez-de{-}Vera et~al.(2022)]{RODRIGUEZDEVERA2022106548}
Rodr{\'i}guez-de{-}Vera~JM, Bernab{\'e}~G, Garc{\'i}a~JM, Daniel~S, Gonz{\'a}lez-Carrillo~J. 2022. Left ventricular non-compaction cardiomyopathy automatic diagnosis using a deep learning approach. Computer Methods and Programs in Biomedicine. 214:106548.  \urlprefix\url{https://www.sciencedirect.com/science/article/pii/S0169260721006222}.

\bibitem[Rodr{\'i}guez-de{-}Vera et~al.(2021)]{lvncDL}
Rodr{\'i}guez-de{-}Vera~JM, Gonz{\'a}lez-Carrillo~J, Garc{\'i}a~JM, Bernab{\'e}~G. 2021. Deploying deep learning approaches to left ventricular non-compaction measurement. The Journal of Supercomputing{ }:--.

\bibitem[Ronneberger et~al.(2015)]{ronneberger_fischer_brox_2015}
Ronneberger~O, Fischer~P, Brox~T. 2015. U-net: convolutional networks for biomedical image segmentation. Lecture Notes in Computer Science Medical Image Computing and Computer-Assisted Intervention – MICCAI 2015:234–241.{ }.

\bibitem[Sze et~al.(2017)]{Survey-AI}
Sze~V, Chen~YH, Yang~TJ, Emer~JS. 2017. Efficient processing of deep neural networks: A tutorial and survey. Proceedings of the IEEE. 105(12):2295--2329.

\bibitem[{Towbin} et~al.(2015)]{towbin2015lvnc}
{Towbin}~JA, {Lorts}~A, {Jefferies}~JL. 2015. Left ventricular non-compaction cardiomyopathy. The Lancet. 386:813--825.

\bibitem[Udeoji et~al.(2013)]{doi:10.1177/1753944713504639}
Udeoji~DU, Philip~KJ, Morrissey~RP, Phan~A, Schwarz~ER. 2013. Left ventricular noncompaction cardiomyopathy: updated review. Therapeutic Advances in Cardiovascular Disease. 7(5):260--273. PMID: 24132556.

\bibitem[{World Health Organization}(2022)]{CVDsWHO}
{World Health Organization}. 2022. Cardiovascular diseases; [\url{https://www.who.int/health-topics/cardiovascular-diseases}]. Accessed: \today.

\bibitem[Zaid et~al.(2008)]{Zaid-08}
Zaid~AO, Abdessalem~S, Mourali~MS, Farhati~A, Bouallègue~A, Mechmeche~R, Olivier~C. 2008. Coronary angiogram video compression adapted to medical imaging applications. Annual International Conference of the IEEE Engineering in Medicine and Biology Society:410--413.

\end{thebibliography}

\end{document}